\documentclass[aip,jcp,preprint]{revtex4-1}

\usepackage{amsfonts}
\usepackage{graphicx}
\usepackage{amsmath}
\usepackage{gensymb}
\usepackage{color}

\begin{document}

\title{
\begin{center}
Radiative Emission of Polaritons Controlled by Light-Induced Geometric Phase
\end{center}
}

\author{Csaba F\'abri}
\email{ficsaba@staff.elte.hu}
\affiliation{MTA-ELTE Complex Chemical Systems Research Group, P.O. Box 32, H-1518 Budapest 112, Hungary}
\affiliation{Department of Theoretical Physics, University of Debrecen, PO Box 400, H-4002 Debrecen, Hungary}

\author{G\'abor J. Hal\'asz}
\affiliation{Department of Information Technology, University of Debrecen, P.O. Box 400, H-4002 Debrecen, Hungary}

\author{Lorenz S. Cederbaum}
\affiliation{Theoretische Chemie, Physikalisch-Chemisches Institut, Universit\"at Heidelberg, Im Neuenheimer Feld 229, D-69120 Heidelberg, Germany }

\author{\'Agnes Vib\'ok}
\email{vibok@phys.unideb.hu}
\affiliation{Department of Theoretical Physics, University of Debrecen, PO Box 400, H-4002 Debrecen, Hungary}
\affiliation{ELI-ALPS, ELI-HU Non-Profit Ltd, H-6720 Szeged, Dugonics t\'er 13, Hungary}

\begin{abstract}
Polaritons -- hybrid light-matter states formed in cavity -- strongly change the properties of the underlying matter.
In optical or plasmonic nanocavities, polaritons decay by radiative emission of the cavity, which is accessible experimentally.
Due to the interaction of a molecule with the quantized radiation field, polaritons exhibit light-induced conical intersections (LICIs)
which dramatically influence the nuclear dynamics of molecular polaritons. We show that ultrafast radiative emission from
the lower polariton is controlled by the geometric phase imposed by the LICI. This finding provides insight into the process of emission
and, furthermore, allows one to compute these signals by augmenting the Born-Oppenheimer approximation for polaritons
with a geometric phase term.
\end{abstract}

\maketitle

Conical intersections (CIs) are degeneracies between close-lying electronic states of
molecules.\cite{Koppel198459,Yarkony1996985,Baer_2002,Domcke2004,Worth2004127,Baer2006,02WoRo,08SiBlBe}
CIs are ubiquitous in polyatomic molecules which possess a dense manifold of electronic
states and several nuclear degrees of freedom. CIs enable radiationless
transitions which take place between electronic states in the vicinity
of nuclear configurations where the relevant potential energy surfaces (PESs) intersect.
In order to properly describe ultrafast nonadiabatic processes one needs to
invoke the coupled-electronic-state nonadiabatic approach instead
of the so-called single-surface Born--Oppenheimer (BO) scheme
\cite{Born1927} which breaks down in the presence of CIs.

CIs can also emerge when the molecule is exposed
to classical laser light \cite{Sindelka2011} or to the quantized radiation
field of an optical or plasmonic nanocavity.\cite{Szidarovszky20186215,Csehi2019a}
These CIs are termed light-induced conical intersections (LICIs).
In case of quantized light, the confined photonic mode of the cavity can
couple the electronic states of the molecule, which gives rise to
polaritonic states carrying both photonic and excitonic characters.
Similarly to natural CIs, LICIs can also be harnessed to
modify and control different topological, spectroscopic and dynamical properties of molecules. \cite{Halasz2015348,Kowalewski20162050,Feist2018205,Szidarovszky20186215,Fregoni2018,Ribeiro2018242,Csehi2019,Csehi2019a,Ulusoy20198832,Fabri2020234302,Felicetti20208810,Gu20201290,Gu2020a,D1CP00943E,Fabri2021a,Cederbaum2021a,Szidarovszky2021,Badanko2022}
In sharp contrast to natural CIs, the position of the LICI and the strength of the light-induced
nonadiabatic coupling can be controlled by tuning the parameters of the classical or quantized light field.\cite{Sindelka2011,Halasz2015348}
The extremely small mode volumes of plasmonic cavities provide huge coupling strengths
(in the $1-100 ~ \textrm{meV}$ range) to single molecules.\cite{18HuSiSu,22TaCuSu} Strong coupling between one or few molecules
and photons can only be realized in such cavities.\cite{Chikkaraddy2016127}
Nevertheless, plasmonic nanocavities are very bad objects as they have low quality factors and high loss rate.

Inspired by polaritonic molecular clocks,\cite{Silva2020}
we have shown for the four-atomic H$_{2}$CO (formaldehyde) molecule
that the time-resolved ultrafast radiative emission of
the cavity enables one to follow nonadiabatic population transfer
between polaritonic surfaces.\cite{Fbri2022}
This effect can be seen as an unambiguous
(and in principle experimentally accessible) dynamical fingerprint of the LICI.\cite{Fbri2022}
Here, we investigate the LICI-affected ultrafast radiative emission from the
lower polaritonic state and demonstrate the role of the geometric
phase\cite{Ryabinkin2013,Ryabinkin20171785,Henshaw_2017,Izmaylov20165278,Kendrick19972431}
(GP) associated with the LICI.

Nonadiabatic dynamics near a CI can be treated either in the adiabatic or diabatic representation.
The description of dynamics in the vicinity of a CI in the adiabatic
representation reflects GP as a clear consequence
that each real-valued adiabatic electronic wave function changes sign when transported
continuously along a closed loop enclosing the CI.\cite{Mead19792284}
As the total molecular wave function has to be single valued,
we choose, as often done,\cite{Mead19792284} to make the adiabatic electronic wave function complex
by multiplying it with a position-dependent phase factor, which
ensures that the total wave function remains single valued.
This modification of the electronic wave function has a direct effect on nuclear dynamics
even when a single PES is considered. One way to include the GP explicitly in the single-state BO approximation
is the vector potential approach \cite{Mead19792284} which is referred
to as the molecular Aharonov--Bohm effect.\cite{Aharonov1959,AldenMead198023}
In several situations GP is essential to obtain qualitatively 
correct results when problems are considered in the single-state
BO approximation.\cite{Cederbaum20034,Xie20167828,Xie20171902,Xie2017,Xie20181986,Ryabinkin2013,Henshaw_2017}
Here we demonstrate that the BO approximation can substantially mistake
radiative emission from the lower polaritonic state, which can be ascribed to neglecting the GP
in certain situations.

A molecule coupled to a single cavity mode is described by the Hamiltonian
\begin{equation}
	\hat{H}_\textrm{cm} = \hat{H}_0 + \hbar \omega_\textrm{c} \hat{a}^\dag \hat{a} - g \hat{\vec{\mu}} \vec{e} (\hat{a}^\dag + \hat{a}).
   \label{eq:Hcm}
\end{equation}
In eqn \eqref{eq:Hcm} $\hat{H}_0$ refers to the Hamiltonian of the isolated molecule, $\omega_\textrm{c}$ is the cavity angular frequency,
$\hat{a}^\dag$ and $\hat{a}$ denote creation and annihilation operators, $g$ is the coupling strength parameter,
$\hat{\vec{\mu}}$ corresponds to the molecular electric dipole moment operator and $\vec{e}$ is the polarization vector of the cavity field.
In what follows, two molecular electronic states (X and A) will be considered.

Following Ref. \citenum{Silva2020}, the cavity mode is pumped with a laser pulse, which can be described by the Hamiltonian
\begin{equation}
	\hat{H} = \hat{H}_\textrm{cm} + \hat{H}_\textrm{L} ~~~ \textrm{with} ~~~ \hat{H}_\textrm{L} = - \mu_\textrm{c} E(t) (\hat{a}^\dag+\hat{a}).
  \label{eq:fullH}
\end{equation}
In eqn \eqref{eq:fullH} the effective dipole moment of the cavity mode is set to $\mu_\textrm{c} = 1.0 ~ \textrm{au}$
and the laser field has the form $E(t) = E_0 \sin^2(\pi t / T) \cos(\omega_\textrm{L} t)$ for $0 \le t \le T$,
and $E(t)=0$ otherwise.
$E_0$, $T$ and $\omega_\textrm{L}$ are the amplitude, length and carrier angular frequency of the laser pulse.

It is often necessary to account for the finite lifetime of the cavity mode, therefore, we employ the Lindblad equation\cite{20Manzano}
\begin{equation}
\frac{\partial \hat{\rho}}{\partial t} =
		-\textrm{i} [\hat{H},\hat{\rho}] + \gamma_\textrm{c} \hat{a} \hat{\rho} \hat{a}^\dag -
		 \frac{\gamma_\textrm{c}}{2} ( \hat{\rho} \hat{N} + \hat{N} \hat{\rho} )
	\label{eq:Lindblad}
\end{equation}
to describe the quantum dynamics of the coupled cavity-molecule system.
In eqn \eqref{eq:Lindblad} $\hbar=1$ is assumed, $\hat{\rho}$ is the density operator, $\hat{N} = \hat{a}^\dag \hat{a}$ 
denotes the photon number operator and $\gamma_\textrm{c}$ is the cavity decay rate. Throughout this work
we have applied the value $\gamma_\textrm{c} = 10^{-4} ~ \textrm{au}$ which translates to a lifetime of
$1 / \gamma_\textrm{c} =  241.9 ~ \textrm{fs}$. The radiative  emission rate is expressed as $E_\textrm{R} \sim N(t)$
where $N(t) = \textrm{tr}(\hat{\rho}\hat{N})$ is the expectation value of $\hat{N}$.\cite{Silva2020}

Next, the Hamiltonian of eqn \eqref{eq:fullH} is represented in the diabatic basis (direct product of
molecular electronic states and Fock states of the cavity mode). The diabatic Hamiltonian
which includes both nonadiabatic and GP effects is transformed to the adiabatic representation
by diagonalizing the diabatic potential energy matrix. 
Polaritonic PESs are obtained as the eigenvalues of the diabatic potential energy matrix.
As usual, the BO Hamiltonian $\hat{H}^\textrm{BO}$ is defined by omitting nonadiabatic coupling
terms.\cite{Koppel198459,Yarkony1996985,Baer_2002,Domcke2004,Worth2004127,Baer2006}
Note that $\hat{H}^\textrm{BO}$ excludes both nonadiabatic and GP effects.
The latter can be taken into account by the Mead and Truhlar
approach\cite{Mead19792284} which involves the multiplication of the nuclear wave function with
a position-dependent phase factor $\exp(-\textrm{i}\theta)$ introducing
a sign change of the nuclear wave function along closed loops encircling the CI.
The current work focuses on pumping the system to the singly-excited subspace (ground
electronic state with one photon and excited electronic state with zero photon) which
accommodates the lower (LP) and upper (UP) polaritonic states.
Therefore, $\theta$ can be chosen as the transformation angle which parameterizes the
two-by-two diabatic-to-adiabatic transformation in the singly-excited subspace.
These considerations lead to the Hamiltonian
\begin{equation}
	\hat{H}^\textrm{BO}_\textrm{GP} = 
                             \hat{H}^\textrm{BO} + \frac{\textrm{i}}{2}((\nabla \theta)\nabla+\nabla(\nabla \theta)) + 
                             \frac{1}{2} (\nabla \theta)^2
	\label{eq:hadapproxGP}
\end{equation}
where $\hat{H}^\textrm{BO}$ is supplemented by terms describing GP 
effects.\cite{Mead19792284,Ryabinkin2013,Ryabinkin20171785,Henshaw_2017,Izmaylov20165278}
We stress that $\hat{H}^\textrm{BO}$ in eqn \eqref{eq:hadapproxGP} corresponds to either the LP or the UP states
which are the most relevant polaritonic states in our case.
Obviously, $\hat{H}^\textrm{BO}_\textrm{GP}$ includes GP effects, but excludes the possibility of nonadiabatic transitions.
Guo and co-workers investigated interesting properties of natural conical
intersections of isolated molecules along a similar line.\cite{Xie20171902,Xie2017}
A comprehensive description of the necessary theory and the diabatic and adiabatic representations is provided in the SI.

We utilize a two-dimensional vibrational model of H$_2$CO, called the
2D($\nu_2$,$\nu_4$) model.\cite{Fabri2020234302,Fabri2021a,Fbri2022}
The 2D($\nu_2$,$\nu_4$) model treats the two singlet electronic states $\textrm{S}_0 ~ (\tilde{\textrm{X}} ~ ^1\textrm{A}_1)$
and $\textrm{S}_1 ~ (\tilde{\textrm{A}} ~ ^1\textrm{A}_2)$, and the $\nu_2$ (C=O stretch) and $\nu_4$ (out-of-plane bend)
vibrational modes.
As concluded by earlier studies,\cite{Fabri2020234302,Fabri2021a,Fbri2022}
the 2D($\nu_2$,$\nu_4$) model provides a physically correct description of the quantum dynamics of H$_2$CO.
Moreover, as two vibrational degrees of freedom are considered, the 2D($\nu_2$,$\nu_4$) model, in contrast to one-dimensional
descriptions, allows for the formation of LICIs between polaritonic PESs.\cite{Fabri2020234302,Fabri2021a,Gu20201290,Gu2020a}
We refer to the SI for further technical details.

\begin{figure}
\centering
  \includegraphics[scale=0.75]{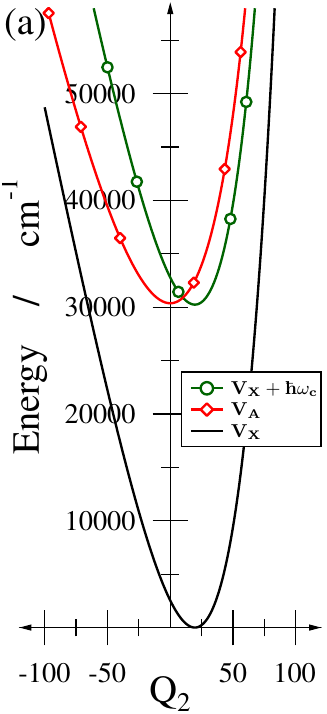}
  \includegraphics[scale=0.9]{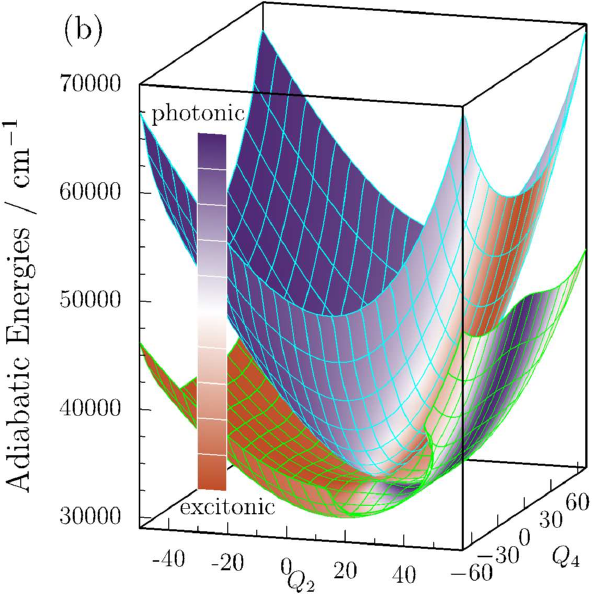}
  \caption{
        (a)
        Diabatic potentials ($V_\textrm{X}$, $V_\textrm{A}$ and $V_\textrm{X}+\hbar \omega_\textrm{c}$)
	as a function of the $Q_2$ (C=O stretch) normal coordinate
	($Q_4 = 0$). The cavity wavenumber equals $\omega_\textrm{c} = 30245.5 ~ \textrm{cm}^{-1}$.
        (b)
        Two-dimensional lower (LP) and upper (UP) polaritonic surfaces
        with $\omega_\textrm{c} = 30245.5 ~ \textrm{cm}^{-1}$ and $g = 0.1 ~ \textrm{au}$ (coupling strength).
	The character of the polaritonic surfaces is indicated by a purple-orange colormap (purple: photonic, orange: excitonic).}
  \label{fig:pes}
\end{figure}

\begin{figure}
\centering
\includegraphics[scale=0.8]{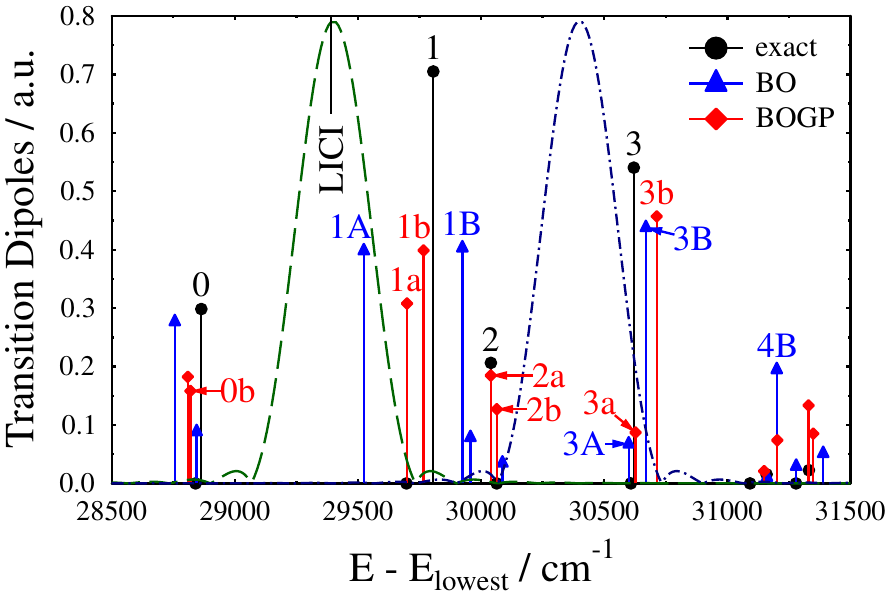}
\caption{
        Absolute values of transition dipoles between selected cavity-molecule eigenstates and the lowest-energy eigenstate 
        for three different models (exact, Born--Oppenheimer (BO) and BO with geometric phase (BOGP)).
        Fourier transforms (absolute value) of the $200 ~ \textrm{fs}$ laser pulses used to initiate the dynamics are also shown
        (carrier wavenumbers:  $\omega_\textrm{L} = 29400 ~ \textrm{cm}^{-1}$ (left curve, dashed line) and 
        $\omega_\textrm{L} = 30400 ~ \textrm{cm}^{-1}$ (right curve, dash-dotted line)).
        Energy levels of selected eigenstates are referenced to the lowest energy level ($E-E_\textrm{lowest}$).
        The energetic position of the light-induced conical intersection (LICI), explicitly marked in the figure, is
        $29390.5 ~ \textrm{cm}^{-1}$ referenced to $E_\textrm{lowest}$.
        The cavity wavenumber and coupling strength are 
        $\omega_\textrm{c} = 30245.5 ~ \textrm{cm}^{-1}$ and $g = 0.1 ~ \textrm{au}$, respectively.}
\label{fig:dipole}
\end{figure}

If the cavity wavenumber and coupling strength are chosen as
$\omega_\textrm{c} = 30245.5 ~ \textrm{cm}^{-1}$ and $g = 0.1 ~ \textrm{au}$,
the LICI is located at $Q_2 = 10.05$ and $Q_4 = 0$ ($Q_2$ and $Q_4$ are dimensionless normal coordinates
of the modes $\nu_2$ and $\nu_4$) at an energy corresponding to $30897.6 ~ \textrm{cm}^{-1}$
above the minimum of the ground-state polaritonic PES (or $29390.5 ~ \textrm{cm}^{-1}$ referenced to the lowest energy level).
The corresponding diabatic and adiabatic LP and UP PESs are provided in Fig. \ref{fig:pes}
where characters (photonic or excitonic) of the LP and UP PESs are indicated by a purple-orange colormap.
It is also visible in Fig. \ref{fig:pes} that the LP and UP PESs form a LICI whose
topological properties have been investigated in Ref. \citenum{Badanko2022}.
Fig. \ref{fig:dipole} shows absolute values of the transition dipole moments between the lowest-energy eigenstate 
(initial state) and selected eigenstates of the cavity-molecule system for the
three models investigated (exact, Born--Oppenheimer (BO) and BO with geometric phase (BOGP)).
Fourier transforms of the laser pulses that are used to transfer population to the LP state are also given in Fig. \ref{fig:dipole}.
Selected energy levels together with eigenstate labels shown in Fig. \ref{fig:dipole} are provided in Table I of the SI.
The corresponding eigenstates can be assigned to the LP PES either exactly (BO and BOGP models) or dominantly (exact model).
Photonic part populations, defined as the integral of the LP probability density over the photonic region of the LP PES,
are also specified in the SI.

\begin{figure}
\centering
\includegraphics[scale=0.5]{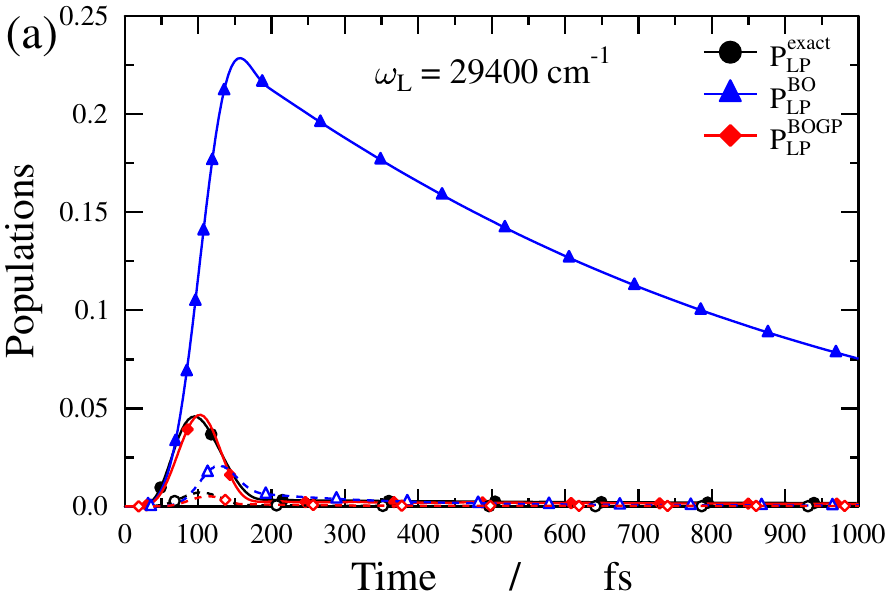}
\includegraphics[scale=0.5]{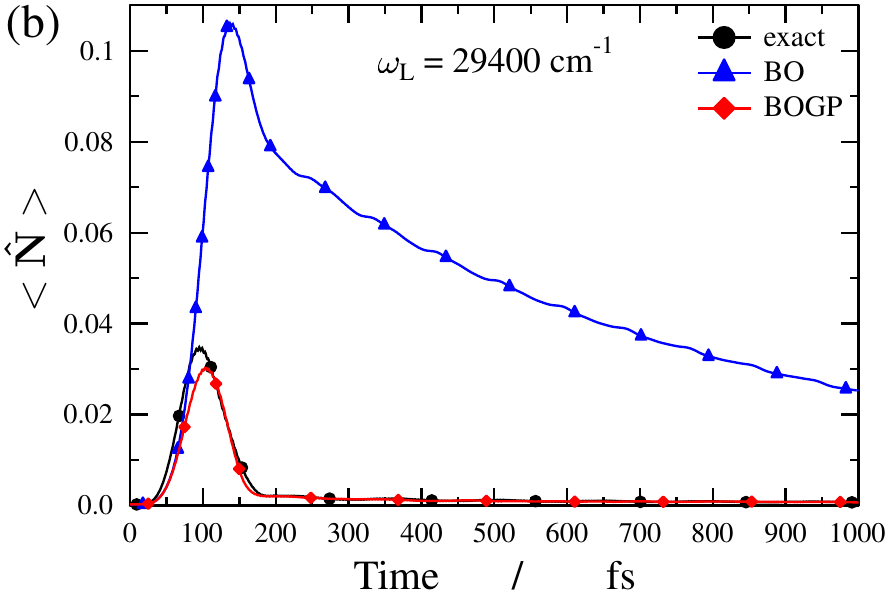}
\includegraphics[scale=0.5]{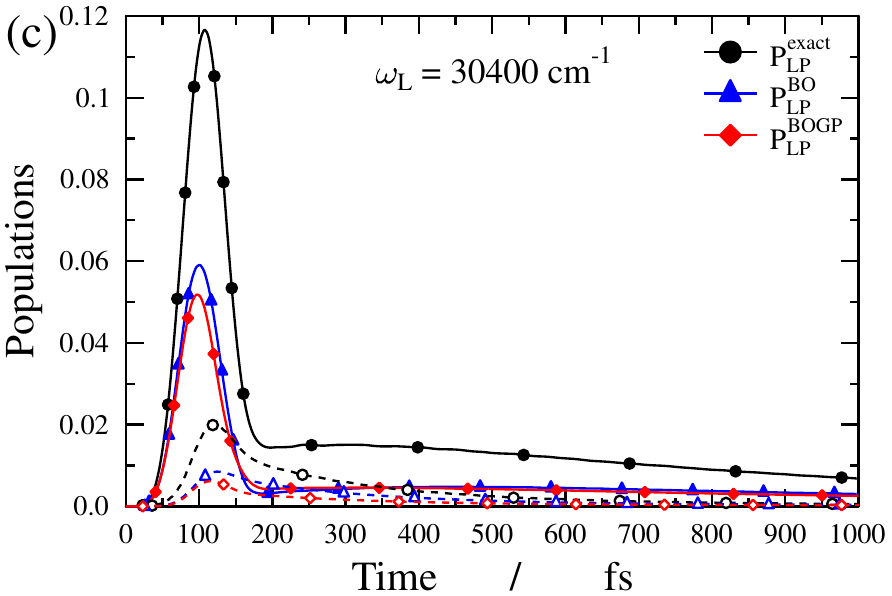}
\includegraphics[scale=0.5]{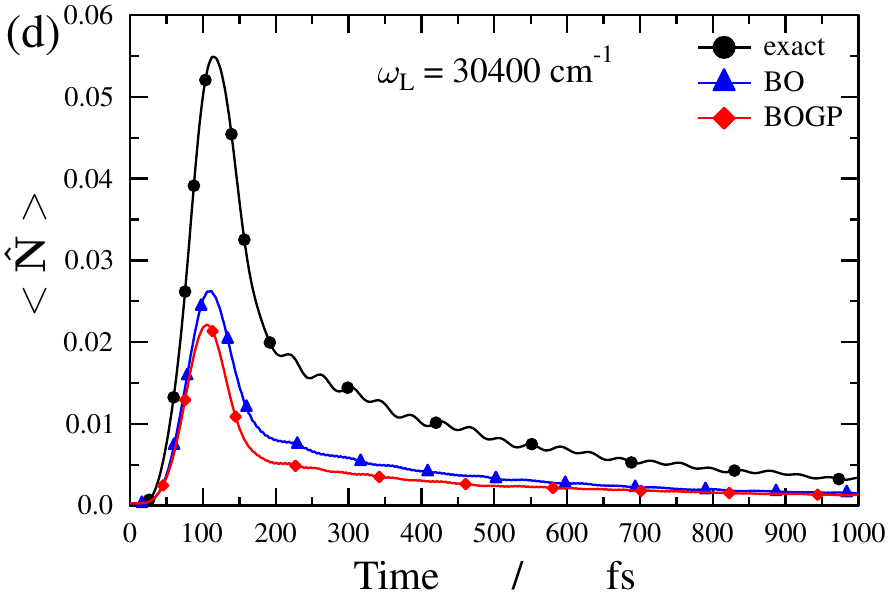}
\caption{
		(a) 
		Population of the lower polaritonic (LP) state for three different models
		(exact, Born--Oppenheimer (BO) and BO with geometric phase (BOGP)) during and after excitation
		with a $T = 200 ~ \textrm{fs}$ laser pulse (carrier wavenumber: $\omega_\textrm{L} = 29400 ~ \textrm{cm}^{-1}$).
		The cavity wavenumber and coupling strength are $\omega_\textrm{c} = 30245.5 ~ \textrm{cm}^{-1}$ and $g = 0.1 ~ \textrm{au}$.
		(b)
		Ultrafast emission signals for the three different models with the parameters of panel a.
		The exact emission is significantly overestimated by the BO model, while
		the BOGP model shows an excellent agreement with the exact results.
		(c-d)
		Same as for panels a-b with $\omega_\textrm{L} = 30400 ~ \textrm{cm}^{-1}$.
		In contrast to panels a-b, the exact emission is underestimated by the BO model and
		inclusion of the GP does not improve the BO model.
		}
\label{fig:gpresults}
\end{figure}

In Fig. \ref{fig:gpresults} population of the LP state and the ultrafast emission signal for the exact, BO and BOGP models are presented.
In panels a and b of Fig. \ref{fig:gpresults}
the cavity mode is pumped with a laser pulse of $\omega_\textrm{L} = 29400 ~ \textrm{cm}^{-1}$, $T = 200 ~ \textrm{fs}$ and
$E_0 = 0.001 ~ \textrm{au}$ (corresponding to a field intensity of $I = 3.51 \cdot 10^{10} ~ \textrm{W} / \textrm{cm}^{2}$),
which leads to population transfer from the initially populated lowest-energy eigenstate to the LP PES with high selectivity.
The value $\omega_\textrm{L} = 29400 ~ \textrm{cm}^{-1}$ essentially coincides with the LICI energy
of $29390.5 ~ \textrm{cm}^{-1}$ (referenced to the lowest energy level).
It is conspicuous in Fig. \ref{fig:gpresults} that the maximal LP population for the BO model is approximately
five times as large as the corresponding exact and BOGP values. In each model the ultrafast emission signal
(proportional to the expectation value of the photon number operator $\hat{N}$) follows the shape of the corresponding LP population curve.
Moreover, while the exact LP population and emission values are significantly overestimated by the BO model,
the exact results agree well with their BOGP counterparts.

In order to understand these observations, populations and probability densities of the relevant eigenstates
are analyzed for the three different models (see Figs. 2 and 3 in the SI).
For the BO model a single eigenstate (denoted by $1\textrm{A}$) is populated dominantly with a maximal population of
0.23  which is roughly five times as large as the maximal population of the dominant
exact ($1$) and BOGP ($1\textrm{a}$ and $1\textrm{b}$) eigenstates.
In addition, maximal populations of the dominant exact and BOGP eigenstates sum up to a nearly identical value (about $0.04$).
In all models, the dominantly-populated eigenstates lie around the
energetic position of the LICI (see Table I of the SI for energy level values).
Of course, in each case several other eigenstates are also populated to some extent, but the dominant populations
can be attributed to the few eigenstates mentioned.

The previous analysis of populations together with the fact that
photonic part populations of the dominant eigenstates (see Table I of the SI) do not differ significantly from each other
explain the origin of both the overestimation of the BO model and the good agreement between the exact and BOGP results.
The conclusions drawn about eigenstate populations are also supported by Fig. \ref{fig:dipole} where transition dipole moments and
energetic positions of relevant eigenstates are highlighted.
One can observe in Fig. \ref{fig:dipole} that the energy of the BO eigenstate labeled $1\textrm{A}$ lies closer to the
center of the Fourier transform of the pulse with $\omega_\textrm{L} = 29400 ~ \textrm{cm}^{-1}$ than
the dominant BOGP ($1\textrm{a}$ and $1\textrm{b}$) and exact ($1$) eigenstates, which explains
why the BO eigenstate $1\textrm{A}$ can acquire substantially higher maximal population than the dominant exact and BOGP eigenstates.
Another interesting observation is that while eigenstates $1$, $1\textrm{a}$ and $1\textrm{b}$ tend to avoid the LICI,
eigenstate $1\textrm{A}$ has its maximal amplitude in the vicinity of the LICI (see Fig. 3 of the SI).
This finding justifies why the BO results are qualitatively different from the exact and BOGP ones
and further supports the excellent agreement between the exact and BOGP models in this particular case.

A significantly different situation is presented panels c and d of Fig. \ref{fig:gpresults} where the cavity mode is pumped
with a laser pulse of $\omega_\textrm{L} = 30400 ~ \textrm{cm}^{-1}$, $T = 200 ~ \textrm{fs}$ and $E_0 = 0.001 ~ \textrm{au}$
and the dominant cavity-molecule eigenstates populated by the laser pulse
 are well above the LICI energy (see Table I of the SI for energy levels).
Fig. \ref{fig:gpresults} reveals that the exact model gives rise to the largest LP population and emission values in this case.
Since the GP correction slightly reduces the BO LP population and emission,
the BOGP results get even further from their exact counterparts.
As a consequence, the BOGP model is not able to approximate the exact description.
Careful analysis of the populations and photonic part populations of the relevant
eigenstates (see Fig. 2 in the SI) serves as an
unequivocal explanation for the emission results obtained with the three different models.
Here, the dominant eigenstates are $3$ (exact), $3\textrm{B}$ (BO) and $3\textrm{b}$ (BOGP)
(see Fig. 4 of the SI for probability densities)
which reach maximal population of $0.10$, $0.04$ and $0.03$, respectively, while possessing almost identical
photonic part populations. The maximal population values can be rationalized by comparing the energetic positions
of the eigenstates $3$, $3\textrm{B}$ and $3\textrm{b}$ to the Fourier transform of the laser pulse with
$\omega_\textrm{L} = 30400 ~ \textrm{cm}^{-1}$ in Fig. \ref{fig:dipole}.
This finding helps interpret the shape of the LP population and emission curves.

We note that similar effects can be observed for other cavity setups.
If the cavity wavenumber and coupling strength equal $\omega_{c} =29957.2 ~ \textrm{cm}^{-1}$
and $g = 0.1 ~ \textrm{au}$, respectively (see Fig. 5 of the SI), the LICI is located at $Q_2= 8.84$ and 
$Q_4 = 0$ at an energy corresponding to $30776.1 ~ \textrm{cm}^{-1}$
above the minimum of the ground-state polaritonic PES
(or $29269.0 ~ \textrm{cm}^{-1}$ referenced to the lowest energy level).
The cavity mode is pumped with two different laser pulses of
$\omega_\textrm{L} = 29200 ~ \textrm{cm}^{-1}$ and $\omega_\textrm{L} = 30300 ~ \textrm{cm}^{-1}$,
both with $T = 200 ~ \textrm{fs}$ and $E_0 = 0.001 ~ \textrm{au}$.
In the first case GP terms are able to correct the BO model. Thus,
the BOGP results agree with the exact ones well, while in the
second case the GP fails to improve upon the BO model. We believe that a remarkable role
is played by the energetic position of the LICI and those of the eigenstates
which are dominantly populated by the pump pulse.
If the relevant energy levels are well above the LICI, we can expect breakdown of the BOGP model
which, in turn, provides very good results around the LICI.

By means of accurate quantum-dynamical calculations, we have simulated the ultrafast radiative emission
signal from the LP surface of the polyatomic H$_2$CO molecule placed in a plasmonic nanocavity.
We have shown that in the presence of a light-induced conical intersection,
which is common if the cavity mode couples two electronic states of a polyatomic molecule, the BO model breaks down.
However, if the LP surface is populated around the LICI, the BO approximation extended 
with geometric-phase terms (BOGP model) can accurately reproduce the
exact emission signal. In other words, the multistate diabatic representation can be
well approximated by the BOGP model excluding nonadiabatic transitions.
However, this is not always the case as highlighted in this work.
Well above the LICI energy, the BOGP model fails to provide appropriate results.
The current study clearly demonstrates that a remarkable role is played by the
energy of the LICI and those of the eigenstates which are
dominantly populated by the pump pulse. We stress that if relevant energy levels lie
significantly above the LICI in energy, the BOGP model breaks down and it is unable to 
provide an alternative to the exact diabatic representation.

\begin{acknowledgments}
The authors are grateful to NKFIH for financial support (Grant K128396).
\end{acknowledgments}

\bibliography{GP-paper_arxiv}

\newpage

\begin{center}
{\Large Supporting Information}
\end{center}

\section{Theoretical considerations}

The Hamiltonian of a molecule coupled to a single cavity mode has the form
\begin{equation}
	\hat{H}_\textrm{cm} = \hat{H}_0 + \hbar \omega_\textrm{c} \hat{a}^\dag \hat{a} - g \hat{\vec{\mu}} \vec{e} (\hat{a}^\dag + \hat{a}).
   \label{eq:HcmSI}
\end{equation}
We refer to the manuscript regarding the notations used in Eq. \eqref{eq:HcmSI}.
Note that the quadratic dipole self-energy term\cite{20ScRuRo,20MaMoHu,20TaMaZh,Triana2021,22FrGaFe} is omitted
in Eq. \eqref{eq:HcmSI} as it is expected to have negligible effects for the cases investigated in this work.
If two molecular electronic states (X and A) are considered, the Hamiltonian of Eq. \eqref{eq:HcmSI} can be recast
in the direct product basis of the electronic states ($| \textrm{X} \rangle$ and $| \textrm{A} \rangle$) 
and Fock states of the cavity mode ($| n \rangle$ with $n=0,1,2,\dots$) as
\begin{equation}
    \resizebox{0.9\textwidth}{!}{$\hat{H}_\textrm{cm}  = 
         \begin{bmatrix}
            \hat{T} + V_\textrm{X} & 0 & 0 & W_1 & 0 & 0 & \dots \\
            0 & \hat{T} + V_\textrm{A} & W_1 & 0 & 0 & 0 & \dots \\
            0 & W_1 & \hat{T} + V_\textrm{X} + \hbar\omega_\textrm{c} & 0 & 0 & W_2 & \dots \\
            W_1 & 0 & 0 &\hat{T} + V_\textrm{A} + \hbar\omega_\textrm{c} & W_2 & 0 & \dots \\
            0 & 0 & 0 & W_2 &\hat{T} + V_\textrm{X} + 2\hbar\omega_\textrm{c} & 0 & \dots \\
            0 & 0 & W_2 & 0 & 0 &\hat{T} + V_\textrm{A} + 2\hbar\omega_\textrm{c} & \dots \\
            \vdots & \vdots & \vdots & \vdots & \vdots & \vdots & \ddots 
        \end{bmatrix}$}
    \label{eq:cavity_H}
\end{equation}
where $\hat{T}$ denotes the kinetic energy operator, while $V_\textrm{X}$ and $V_\textrm{A}$
are the ground-state and excited-state potential energy surfaces (PESs). The cavity-molecule
coupling is described by the operator $W_n = -g \sqrt{n} \vec{d} \vec{e}$
where $\vec{d}$ is the molecular transition dipole moment vector.
It is important to note that terms pertaining to the X and A permanent dipole moments
are neglected in Eq. \eqref{eq:cavity_H}.

The interaction of the cavity mode with a laser pulse is described by the Hamiltonian
\begin{equation}
	\hat{H}_\textrm{L}= - \mu_\textrm{c} E(t) (\hat{a}^\dag+\hat{a})
  \label{eq:W}
\end{equation}
which gives rise to the total Hamiltonian
\begin{equation}
	\hat{H} = \hat{H}_\textrm{cm} + \hat{H}_\textrm{L}.
\end{equation}
All previous equations correspond to the diabatic representation. The adiabatic representation
is defined by diagonalizing the potential energy part ($V$) of the Hamiltonian in Eq. \eqref{eq:cavity_H},
\begin{equation}
	V^\textrm{ad} = U^\textrm{T} V U
\end{equation}
where $V^\textrm{ad}$ contains the polaritonic PESs on its diagonal.
Accordingly, the Hamiltonian in the adiabatic representation equals
\begin{equation}
	\hat{H}^\textrm{ad} = U^\textrm{T} \hat{H} U = U^\textrm{T} \hat{T} U + V^\textrm{ad} + U^\textrm{T} \hat{H}_\textrm{L} U.
\end{equation}
The Born--Oppenheimer (BO) approximation is defined by neglecting the kinetic coupling terms in $\hat{H}^\textrm{ad}$
(in other words, the approximation $U^\textrm{T} \hat{T} U \approx \hat{T}$ is made), that is,
\begin{equation}
	\hat{H}^\textrm{BO} = \hat{T} + V^\textrm{ad} + U^\textrm{T} \hat{H}_\textrm{L} U.
	\label{eq:hadapprox}
\end{equation}

We stress that here the terms diabatic and adiabatic refer to the coupled cavity-molecule system.
Of course, in the field-free case, the electronic states $| \textrm{X} \rangle$ and $| \textrm{A} \rangle$ are adiabatic
electronic states. However, for a molecule coupled to the cavity mode, the light-matter interaction
terms appear in the potential energy part of Eq. \eqref{eq:cavity_H}. Therefore, $\hat{H}_\textrm{cm}$ corresponds
to the diabatic representation and one can move to the adiabatic representation by diagonalizing the potential
energy part of $\hat{H}_\textrm{cm}$.

As a next step, geometric phase (GP) effects are incorporated by taking the similarity-transformed Hamiltonian
\begin{equation}
	\hat{H}^\textrm{BO}_\textrm{GP} = \exp(\textrm{i}\theta) \hat{H}^\textrm{BO} \exp(-\textrm{i}\theta)
	\label{eq:hadapproxGP0}
\end{equation}
where $\exp(-\textrm{i}\theta)$ is a position-dependent phase factor which will enable us
to work with single-valued nuclear wave functions.\cite{Mead19792284,Ryabinkin2013,Ryabinkin20171785,Henshaw_2017,Izmaylov20165278}
As discussed in the manuscript, the coupled cavity-molecule system is pumped to the singly-excited subspace
(ground electronic state with one photon and excited electronic state with zero photon) by a laser pulse.
If the cavity frequency is in near-resonance with the $\textrm{X} \rightarrow \textrm{A}$ electronic transition,
it is a good approximation to separate matrix elements of the potential energy matrix ($V$) of Eq. \eqref{eq:cavity_H} corresponding to
the singly-excited subspace ($V_\textrm{A}$ and $V_\textrm{X} + \hbar\omega_\textrm{c}$)\cite{Vendrell2018,22Cederbaum}
and work with a two-dimensional block of $V$ defined as
\begin{equation}
    V_\textrm{S} = 
    \begin{bmatrix}
        V_\textrm{A} & W_1 \\
        W_1 & V_\textrm{X} + \hbar\omega_\textrm{c}
    \end{bmatrix}.
    \label{eq:vsdef}
\end{equation}
Therefore, in our particular case, $\theta$ is chosen as the angle which parameterizes the two-by-two orthogonal transformation matrix
\begin{equation}
    U = 
    \begin{bmatrix}
        \cos{\theta} & \sin{\theta} \\
        -\sin{\theta} & \cos{\theta}
    \end{bmatrix}
    \label{eq:theta1}
\end{equation}
which diagonalizes $V_\textrm{S}$. Thus, the matrix
\begin{equation}
    U^\textrm{T} V_\textrm{S} U =
    \begin{bmatrix}
        \cos{\theta} & -\sin{\theta} \\
        \sin{\theta} & \cos{\theta}
    \end{bmatrix}
    \begin{bmatrix}
        V_\textrm{A} & W_1 \\
        W_1 & V_\textrm{X} + \hbar\omega_\textrm{c}
    \end{bmatrix}
    \begin{bmatrix}
        \cos{\theta} & \sin{\theta} \\
        -\sin{\theta} & \cos{\theta}
    \end{bmatrix}
    \label{eq:theta2}
\end{equation}
is diagonal if
\begin{equation}
    \theta = \frac{1}{2} \arctan \left( \frac{2W_1}{V_\textrm{X}+\hbar\omega_\textrm{c}-V_\textrm{A}} \right).
    \label{eq:theta3}
\end{equation}
This procedure is clearly an approximation which will be further investigated in future work.
An alternative way of evaluating the transformation angle $\theta$ would be the
block diagonalization idea proposed in Refs. \onlinecite{93PaCeKo} and \onlinecite{89CeScMe}.
Finally, we stress that the Lindblad equation is solved using the full cavity-molecule Hamiltonian $\hat{H}$
with $n = 0,\dots,n_\textrm{max}$ (see the next section for more information) 
and only the calculation of the angle $\theta$ involves the two-by-two approximation used in Eqs.
\eqref{eq:vsdef}, \eqref{eq:theta1}, \eqref{eq:theta2} and \eqref{eq:theta3}.

Eq. \eqref{eq:hadapproxGP0} can be rearranged by evaluating the action of the kinetic energy operator
on $\exp(-\textrm{i}\theta)$, which yields
\begin{equation}
	\hat{H}^\textrm{BO}_\textrm{GP} = \hat{H}^\textrm{BO} + \textrm{i} (\nabla \theta) \nabla +
                                      \frac{\textrm{i}}{2} (\nabla^2 \theta) + \frac{1}{2} (\nabla \theta)^2.
	\label{eq:hadapproxGP1}
\end{equation}
In the 2D($\nu_2$,$\nu_4$) model (see the next section for further discussion) used in numerical computations,
$\hat{T} = -\frac{1}{2} \left( \frac{\partial^2}{\partial Q_2^2} + \frac{\partial^2}{\partial Q_4^2} \right)$
and $\nabla = \left( \frac{\partial}{\partial Q_2} , \frac{\partial}{\partial Q_4} \right)$.
By substituting the commutator
\begin{equation}
    [ \nabla, \nabla \theta ] = \nabla (\nabla \theta) - (\nabla \theta) \nabla = \nabla^2 \theta
\end{equation}
into the second GP term ($\frac{\textrm{i}}{2} (\nabla^2 \theta)$) one can show that the sum of the first two GP terms becomes
\begin{equation}
    \textrm{i} (\nabla \theta) \nabla + \frac{\textrm{i}}{2} (\nabla^2 \theta) = 
            \frac{\textrm{i}}{2} ((\nabla \theta) \nabla + \nabla (\nabla \theta)).
\end{equation}
This way, $\hat{H}^\textrm{BO}_\textrm{GP}$ can be transformed to a more symmetric form
\begin{equation}
	\hat{H}^\textrm{BO}_\textrm{GP} = \hat{H}^\textrm{BO} + \frac{\textrm{i}}{2}((\nabla \theta)\nabla+\nabla(\nabla \theta)) + 
                             \frac{1}{2} (\nabla \theta)^2
	\label{eq:hadapproxGP2}
\end{equation}
which was employed in numerical computations carried out in this study.

\newpage

\section{Computational model and technical details}
\begin{figure}[hbt!]
   \begin{center}
      \includegraphics[scale=0.4]{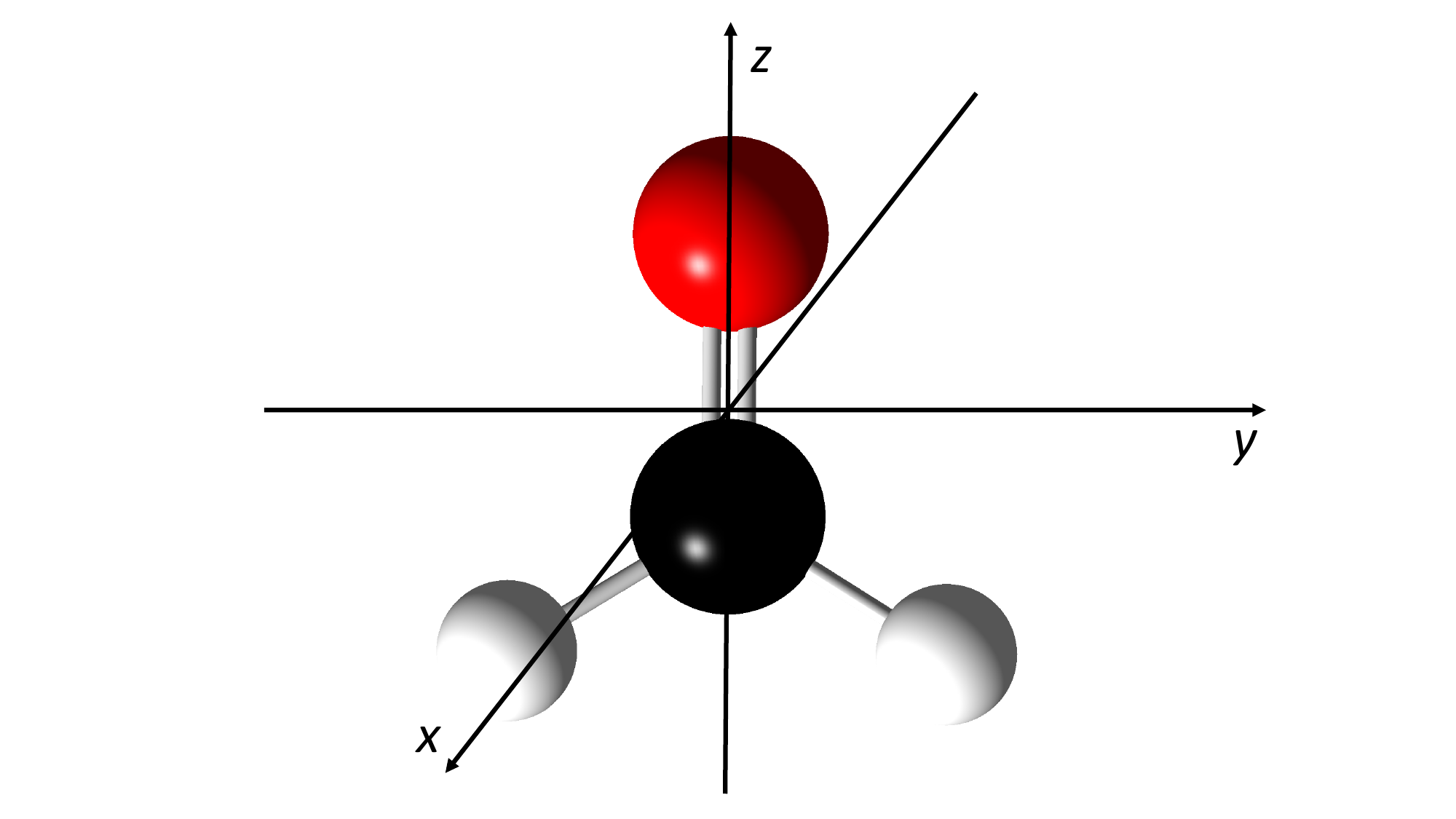}
   \end{center}
   \caption{\label{fig:structure}
        Equilibrium structure of the H$_{2}$CO molecule in the ground electronic state and
        definition of the body-fixed coordinate axes (the equilibrium structure is placed in the $yz$ plane).}
\end{figure}

As already described in previous work,\cite{Fabri2020234302,Fabri2021a,Fbri2022}
the four-atomic formaldehyde (H$_{2}$CO) molecule has a planar equilibrium structure ($C_{2v}$ point-group symmetry)
in the ground electronic state ($\tilde{\textrm{X}} ~ ^1\textrm{A}_1$) and two symmetry-equivalent nonplanar equilibrium structures
($C_{s}$ point-group symmetry) which are connected by a planar transition state structure ($C_{2v}$ point-group symmetry)
in the excited electronic state ($\tilde{\textrm{A}} ~ ^1\textrm{A}_2$).
The ground-state equilibrium structure and definition of the body-fixed coordinate axes are depicted in Figure \ref{fig:structure}.
Out of the six vibrational normal modes of H$_{2}$CO the $\nu_2$ (C=O stretch, $\textrm{A}_1$ symmetry) and
$\nu_4$ (out-of-plane bend, $\textrm{B}_1$ symmetry) vibrational modes are included in the computational model called the 2D($\nu_2$,$\nu_4$) model.
The corresponding anharmonic fundamentals in the ground electronic state (obtained by six-dimensional variational computations)
are $1738.1 ~ \textrm{cm}^{-1}$ ($\nu_2$ mode) and $1147.0 ~ \textrm{cm}^{-1}$ ($\nu_4$ mode).

In order to set up the 2D($\nu_2$,$\nu_4$) model normal coordinates corresponding to the planar transition state structure
of the excited electronic state were evaluated and the four inactive normal coordinates ($Q_1$, $Q_3$, $Q_5$, $Q_6$) were set to zero.
Then, the 2D($\nu_2$,$\nu_4$) potential energy surfaces (PESs) ($V_\textrm{X}$ and $V_\textrm{A}$)
and the transition dipole moment (TDM) surface were computed as a function of the
$Q_2$ and $Q_4$ normal coordinates at the CAM-B3LYP/6-31G* level of theory.
Finally, two-dimensional PES and TDM functions were generated by interpolating the ab initio PES and TDM points.

Due to symmetry, the TDM vanishes at any geometry of $C_{2v}$ symmetry.
Moreover, in the 2D($\nu_2$,$\nu_4$) model, only the body-fixed $y$ component of the TDM can be nonzero and
the TDM is always perpendicular to the permanent dipole moment of both electronic states.
This observation motivates the choice that the cavity field is polarized along the body-fixed $y$ axis in all computations.
Since H$_{2}$CO does not have any first-order nonadiabatic coupling between the X and A electronic states
around its equilibrium geometry, light-induced nonadiabatic effects can be unambiguously distinguished
from natural ones.

The Lindblad equation was solved numerically in the diabatic representation
using the direct product of two-dimensional discrete variable representation basis functions and Fock states of the cavity mode $| n \rangle$ with $n=0,1,2$.
In addition to numerically-exact diabatic computations, Born--Oppenheimer (BO) computations
were carried out without (BO model) or with the GP terms (BOGP model). In both cases the potential energy part of the diabatic Hamiltonian
was diagonalized at each two-dimensional grid point to obtain polaritonic PESs. The Lindblad equation
was then transformed to the adiabatic representation, nonadiabatic coupling terms were omitted and the resulting
equations were solved numerically using the same two-dimensional discrete variable representation basis for each polaritonic PES.

\clearpage

\section{Supplemental data and figures}

\begin{figure}[h]
\includegraphics[scale=0.525]{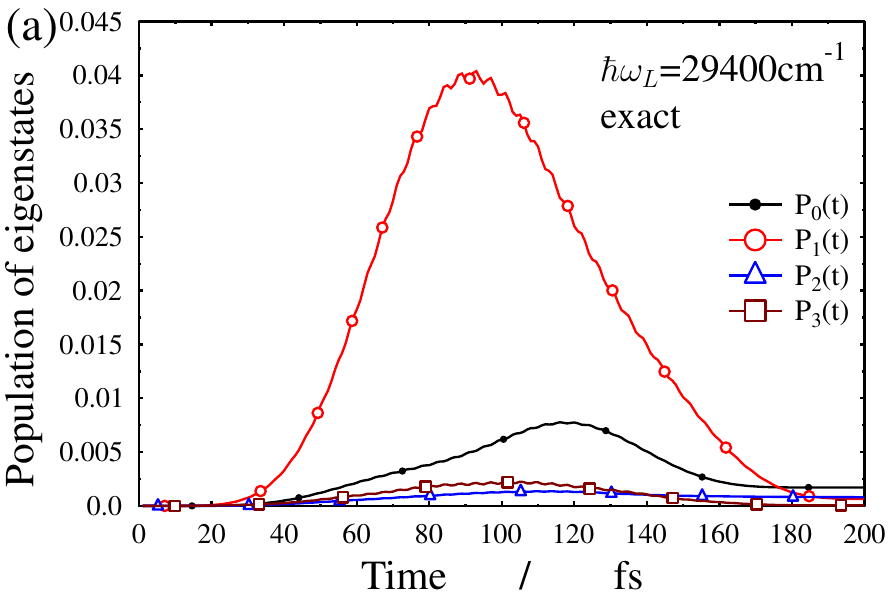}
\includegraphics[scale=0.525]{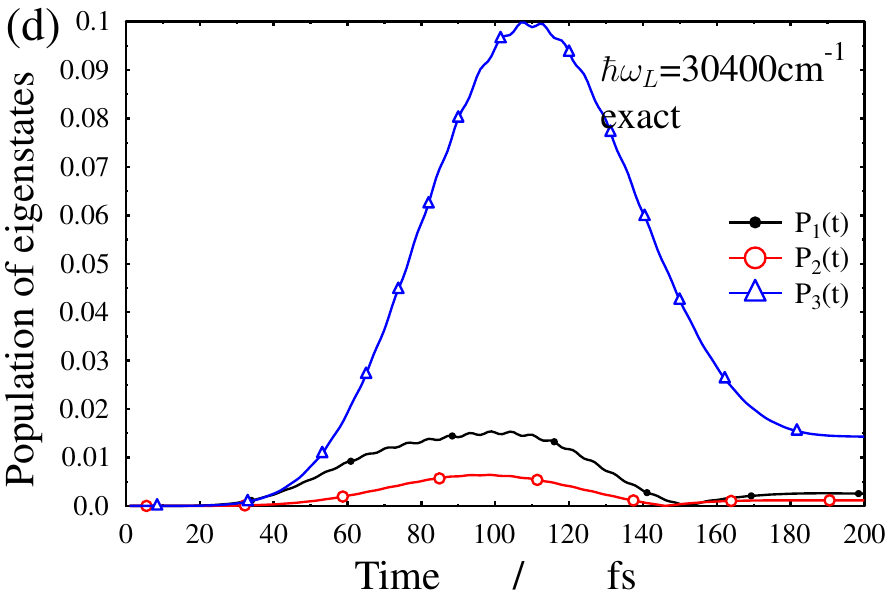}
\includegraphics[scale=0.525]{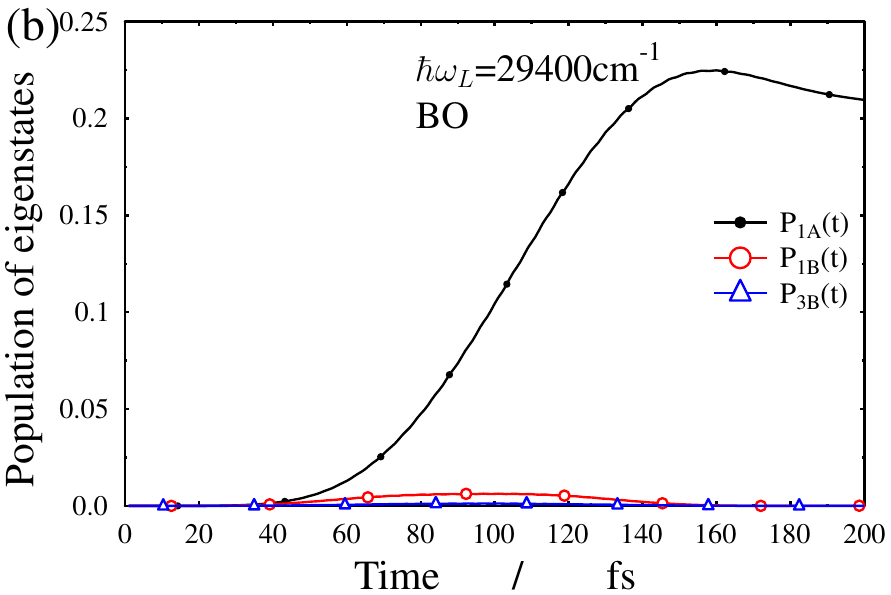}
\includegraphics[scale=0.525]{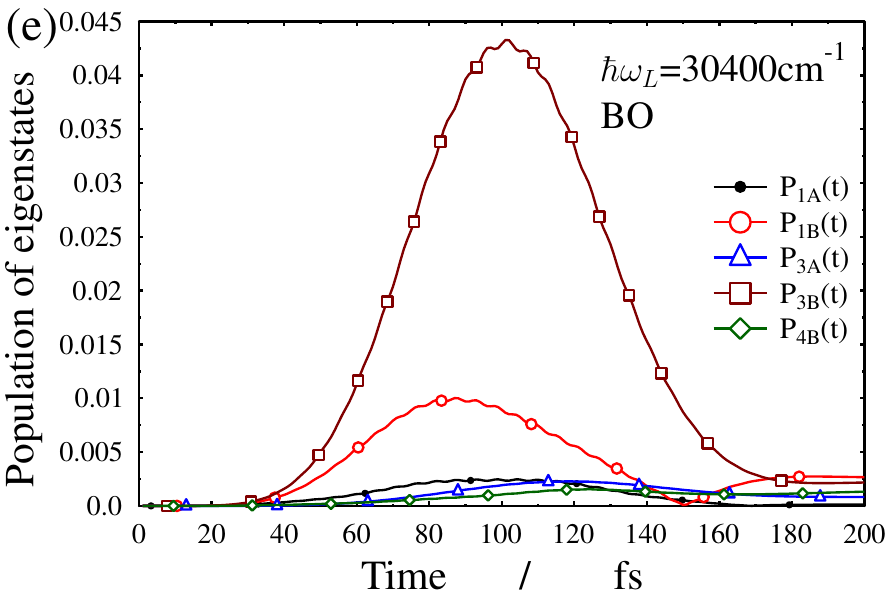}
\includegraphics[scale=0.525]{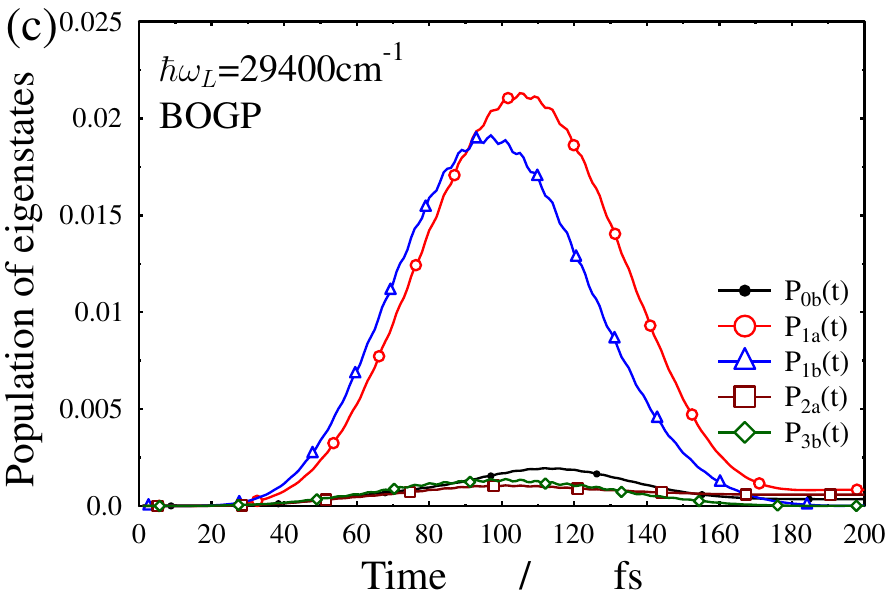}
\includegraphics[scale=0.525]{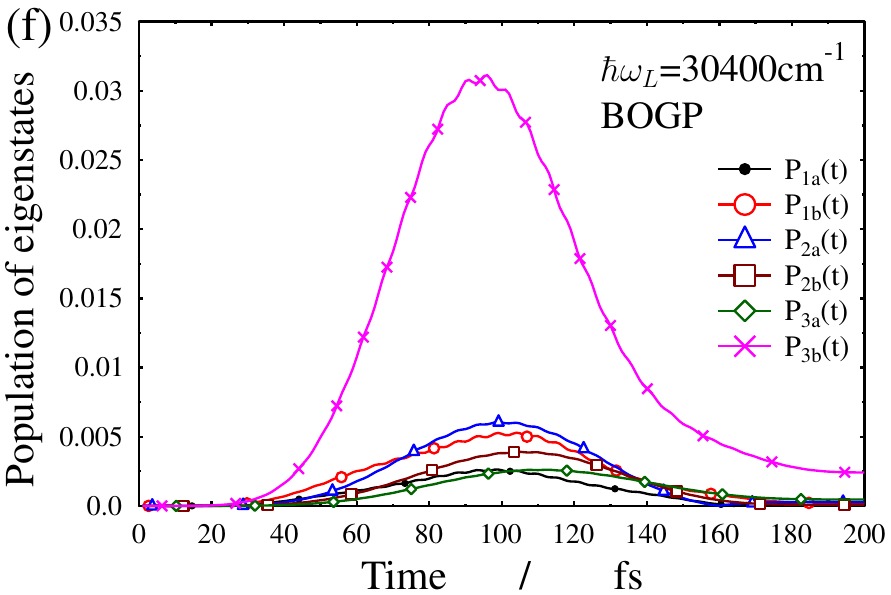}
\caption{
        (a-c)
        Populations of relevant eigenstates of the coupled cavity-molecule system
        for the three different models investigated (exact, Born-Oppenheimer (BO) and BO with geometric phase (BOGP)).
        Populations are shown during excitation with a $200 ~ \textrm{fs}$ laser pulse
        (carrier wavenumber: $\omega_\textrm{L} = 29400 ~ \textrm{cm}^{-1}$).
        (d-f)
        Same as for panels a-c with $\omega_\textrm{L} = 30400 ~ \textrm{cm}^{-1}$.
        Eigenstate labels indicated in the panels are defined in Table \ref{tbl:levels}.
        The cavity parameters are $\omega_{c} = 30245.5 ~ \textrm{cm}^{-1}$ and
        $g = 0.1 ~ \textrm{au}$ for all panels.
        }
\label{fig:populations}
\end{figure}

\begin{figure}[h]
\includegraphics[scale=0.52]{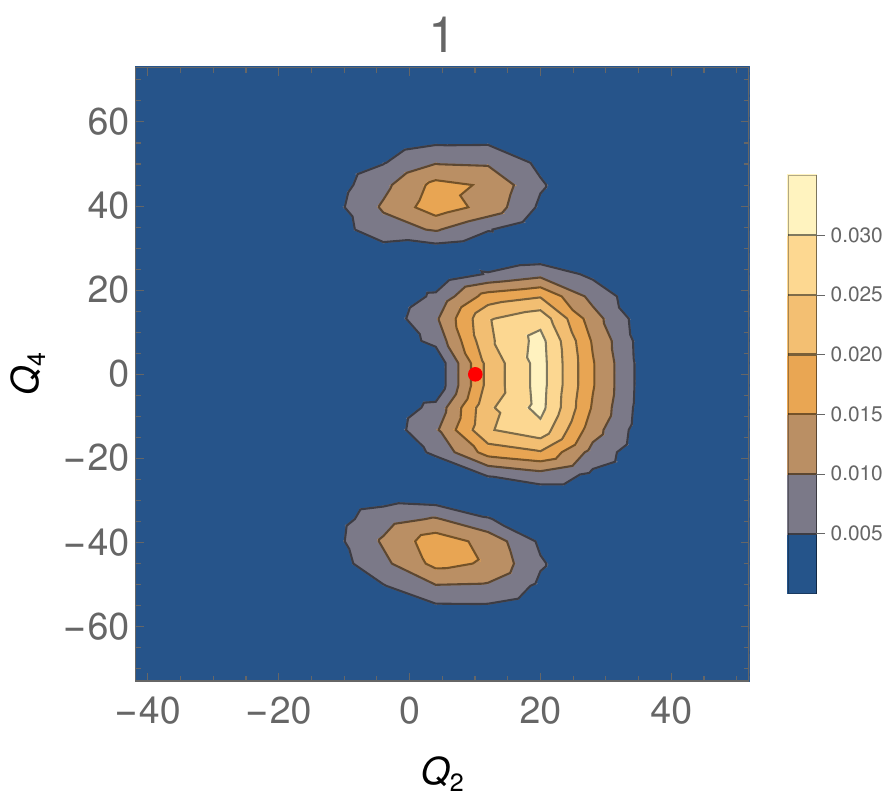}
\includegraphics[scale=0.52]{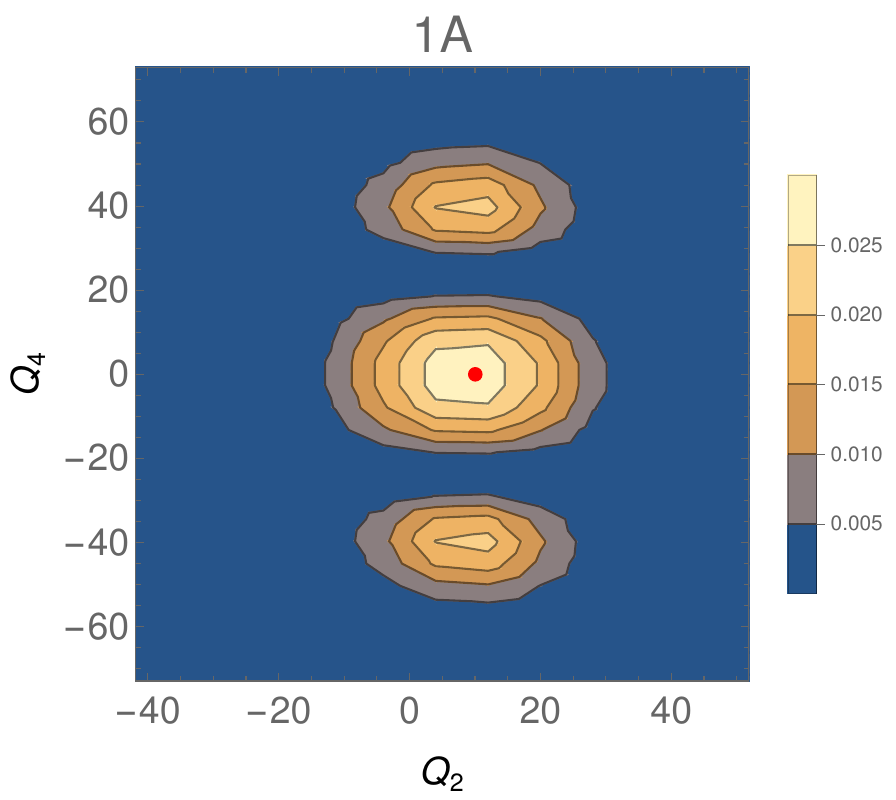}
\includegraphics[scale=0.52]{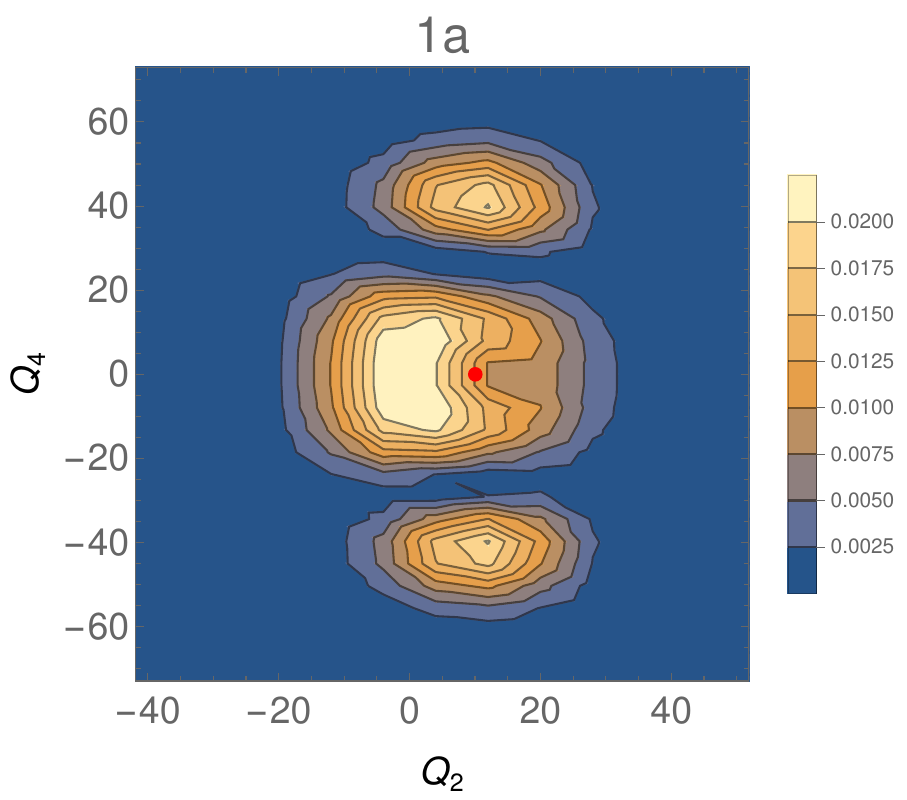}
\includegraphics[scale=0.52]{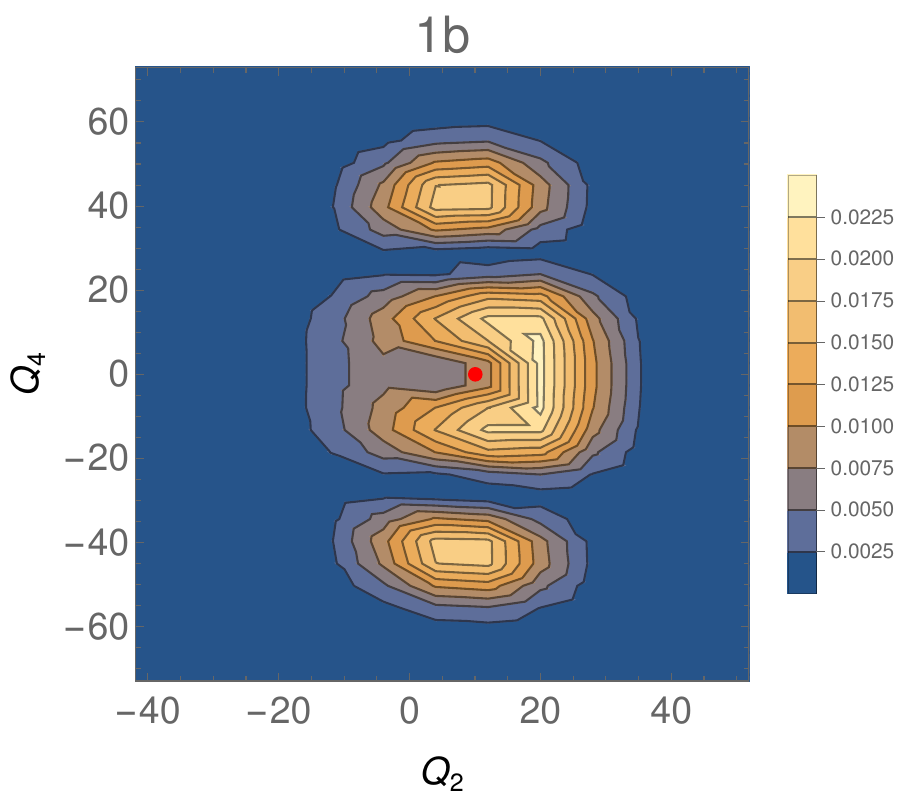}
\caption{\label{fig:pdmgphelps}
		Probability density figures for selected eigenstates of the coupled cavity-molecule system
		(exact: $1$, Born--Oppenheimer (BO): $1\textrm{A}$, BO with geometric phase (BOGP): $1\textrm{a}$ and $1\textrm{b}$,
		see Table \ref{tbl:levels} for more information). $Q_2$ and $Q_4$ are dimensionless
		normal coordinates of the modes $\nu_2$ and $\nu_4$. The cavity wavenumber and coupling strength are 
		$\omega_\textrm{c} = 30245.5 ~ \textrm{cm}^{-1}$ and $g = 0.1 ~ \textrm{au}$, respectively.
		The red dot indicates the position of the LICI at $Q_2 = 10.05$ and $Q_4 = 0$.}
\end{figure}

\begin{figure}[h]
\includegraphics[scale=0.35]{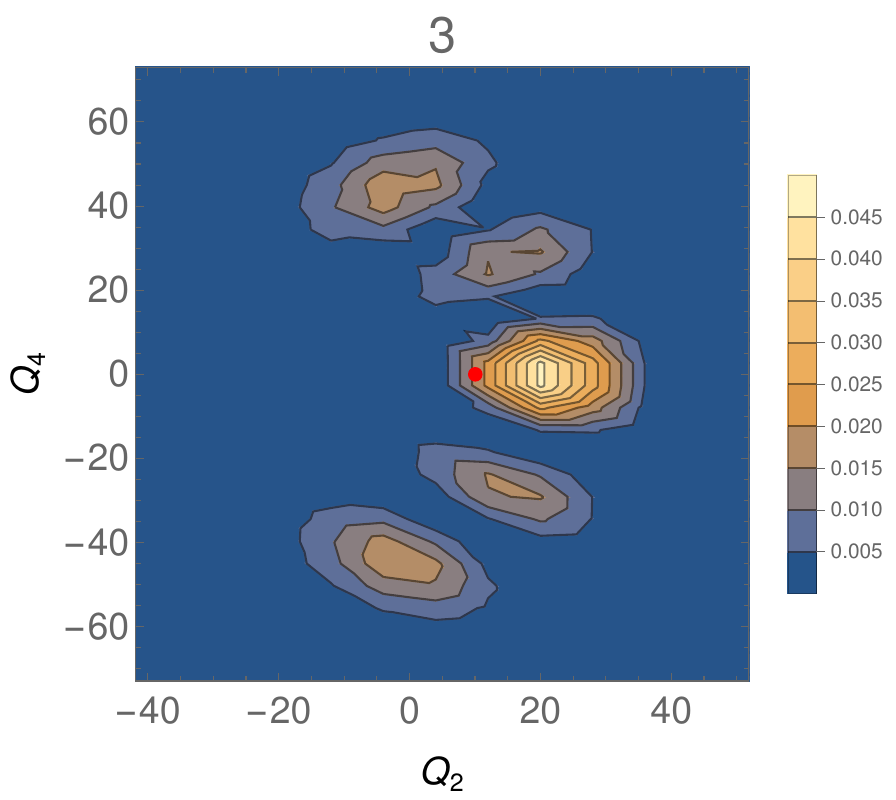}
\includegraphics[scale=0.35]{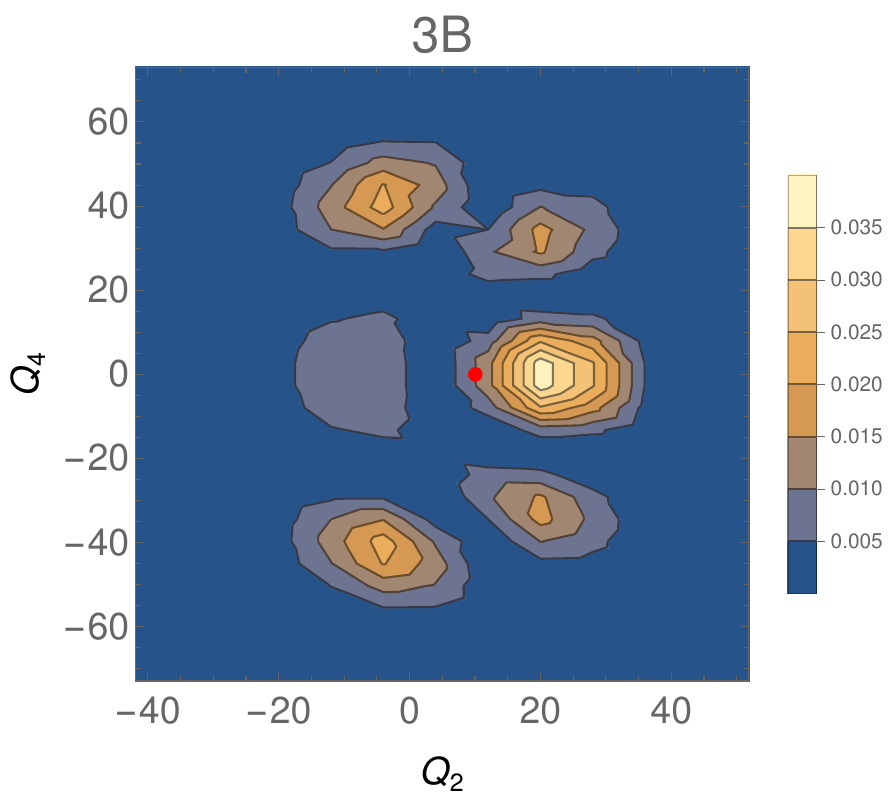}
\includegraphics[scale=0.35]{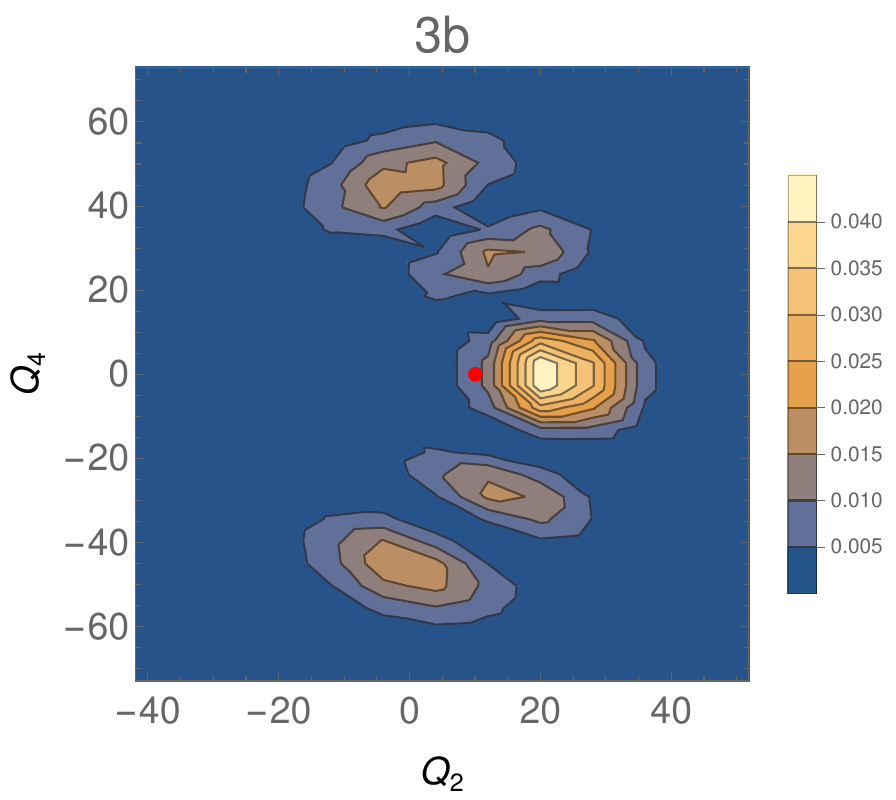}
\caption{\label{fig:pdmgpdoesnothelp}
		Probability density figures for selected eigenstates of the coupled cavity-molecule system
		(exact: $3$, Born--Oppenheimer (BO): $3\textrm{B}$, BO with geometric phase (BOGP): $3\textrm{b}$,
		see Table \ref{tbl:levels} for more information). $Q_2$ and $Q_4$ are dimensionless
		normal coordinates of the modes $\nu_2$ and $\nu_4$. The cavity wavenumber and coupling strength are 
		$\omega_\textrm{c} = 30245.5 ~ \textrm{cm}^{-1}$ and $g = 0.1 ~ \textrm{au}$, respectively.
		The red dot indicates the position of the LICI at $Q_2 = 10.05$ and $Q_4 = 0$.}
\end{figure}

\begin{figure}[h]
\includegraphics[scale=0.54]{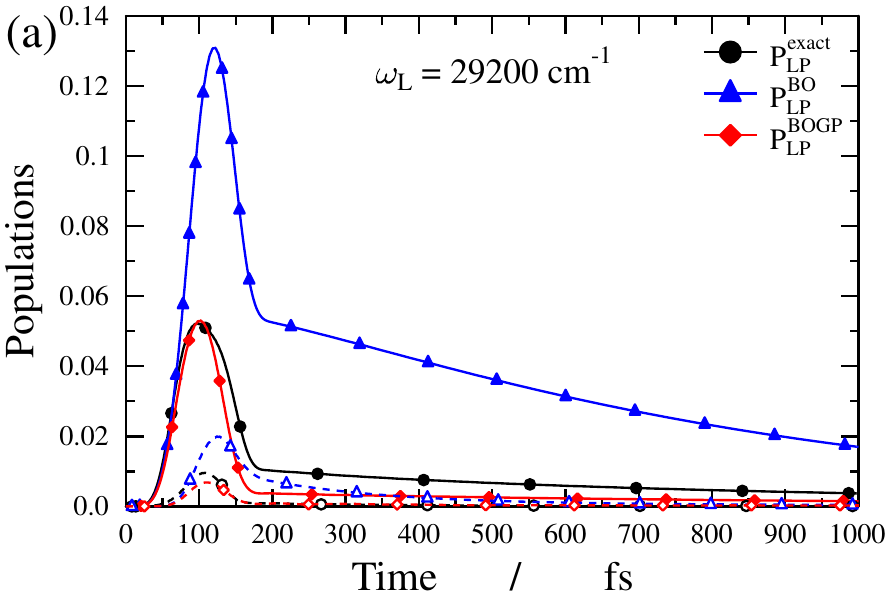}
\includegraphics[scale=0.54]{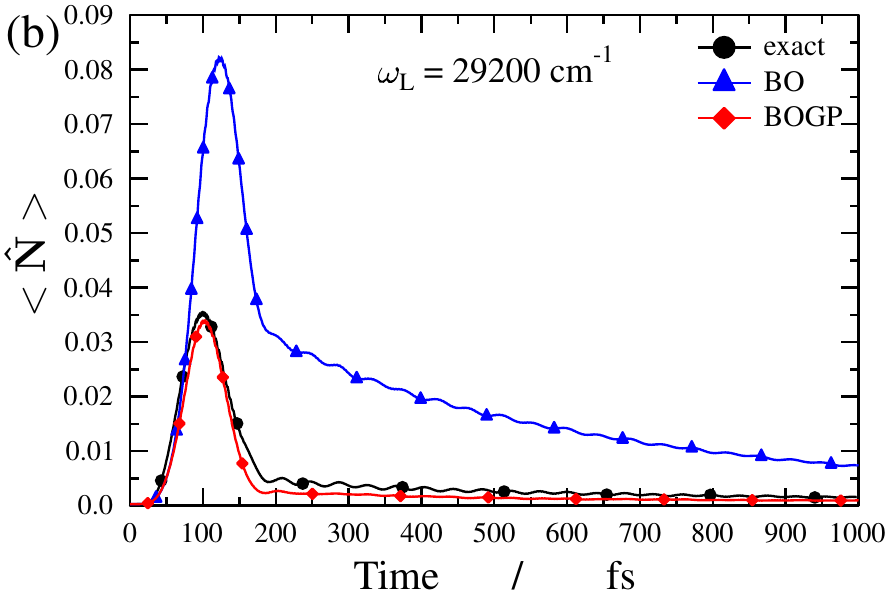}
\includegraphics[scale=0.54]{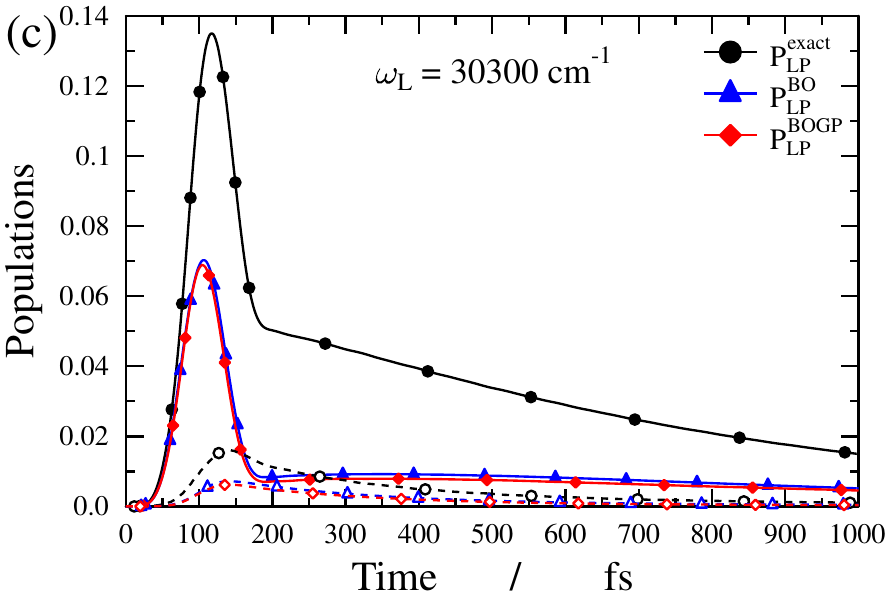}
\includegraphics[scale=0.54]{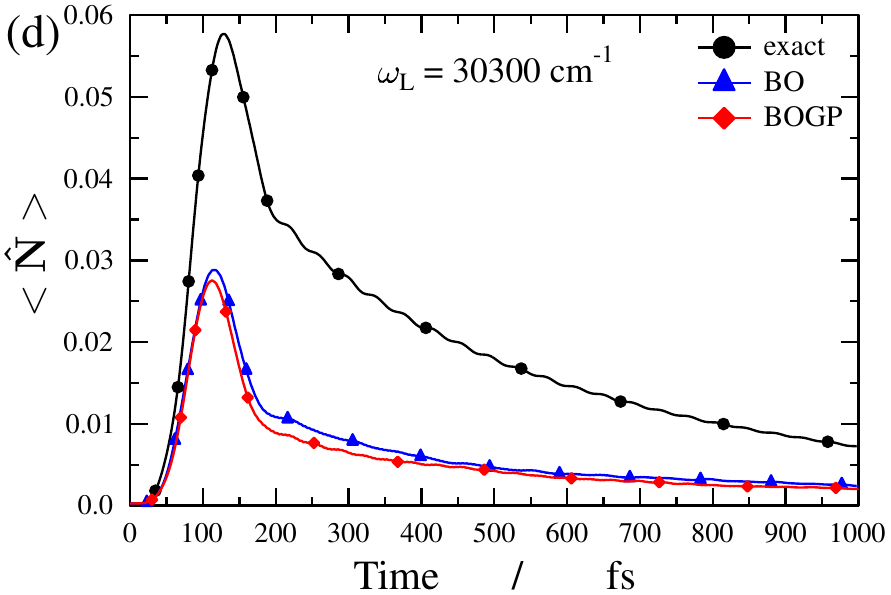}
\caption{\label{fig:gpresultsSI}
		(a) 
		Population of the lower polaritonic (LP) state for the three different models investigated
		(exact, Born--Oppenheimer (BO) and BO with geometric phase (BOGP)) during and after excitation
		with a $200 ~ \textrm{fs}$ laser pulse (carrier wavenumber: $\omega_\textrm{L} = 29200 ~ \textrm{cm}^{-1}$).
		The cavity wavenumber and coupling strength are 
		$\omega_\textrm{c} = 29957.2 ~ \textrm{cm}^{-1}$ and $g = 0.1 ~ \textrm{au}$.
		Populations of polaritonic states higher than LP are negligible (see dashed lines with empty markers).
		(b)
		Ultrafast emission signals for the three different models with the parameters of panel a.
		The emission is proportional to the expectation value of the photon number operator $\hat{N}$.
		The exact emission is significantly overestimated by the BO model, while
		the BOGP model shows an excellent agreement with the exact results.
		(c-d)
		Same as for panels a-b with $\omega_\textrm{L} = 30300 ~ \textrm{cm}^{-1}$.
		In contrast to panels a-b, the exact emission is underestimated by the BO model and
		inclusion of the GP does not improve the BO model.
		}
\end{figure}

Table~\ref{tbl:levels} provides relevant energy levels of the coupled cavity-molecule system together with eigenstate labels and photonic part populations
used in the manuscript (cavity parameters are $\omega_\textrm{c} = 30245.5 ~ \textrm{cm}^{-1}$ and $g = 0.1 ~ \textrm{au}$).
Figure \ref{fig:populations} depicts populations of relevant eigenstates
(exact, Born--Oppenheimer (BO) and BO with geometric phase (BOGP) models)
for the following cavity and laser parameters: $\omega_{c} = 30245.5 ~ \textrm{cm}^{-1}$ and $g = 0.1 ~ \textrm{au}$,
and $\omega_\textrm{L} = 29400 ~ \textrm{cm}^{-1}$ or $\omega_\textrm{L} = 30400 ~ \textrm{cm}^{-1}$,
both with $T = 200 ~ \textrm{fs}$ and $E_0 = 0.001 ~ \textrm{au}$
(corresponding to a field intensity of $I = 3.51 \cdot 10^{10} ~ \textrm{W} / \textrm{cm}^{2}$).
Figures \ref{fig:pdmgphelps} and \ref{fig:pdmgpdoesnothelp} provide probability
density figures for selected eigenstates with $\omega_\textrm{c} = 30245.5 ~ \textrm{cm}^{-1}$ and $g = 0.1 ~ \textrm{au}$
(see Table \ref{tbl:levels} for more information on eigenstate labels).
As exact eigenstates are computed using the diabatic representation, exact eigenstates
are first transformed to the adiabatic representation and LP densities of the resulting states are then evaluated.
Figure \ref{fig:gpresultsSI} shows population and emission figures (exact, BO and BOGP models)
for the cavity parameters  $\omega_{c} =29957.2 ~ \textrm{cm}^{-1}$ and $g = 0.1 ~ \textrm{au}$.
In this case the cavity mode is pumped with the following laser pulses:
$\omega_\textrm{L} = 29200 ~ \textrm{cm}^{-1}$ (panels a-b) and
$\omega_\textrm{L} = 30300 ~ \textrm{cm}^{-1}$ (panels c-d),
both with $T = 200 ~ \textrm{fs}$ and $E_0 = 0.001 ~ \textrm{au}$.

\begin{table}[h]
  \caption{\ Selected energy levels of the coupled cavity-molecule system ($E$ in units of $\textrm{cm}^{-1}$),
             eigenstate labels and photonic part populations  for the three models investigated 
             (exact: labels $0-3$, Born--Oppenheimer (BO): labels $1\textrm{A/B}$, $3\textrm{A/B}$ and $4\textrm{B}$, 
           	 BO with geometric phase (BOGP): labels $0\textrm{b}$, $1\textrm{a/b}$, $2\textrm{a/b}$
           	 and $3\textrm{a/b}$).
           	 Each energy level is referenced to the lowest energy level of the given model
           	 (exact: $E_\textrm{lowest} = 1507.4 ~ \textrm{cm}^{-1}$,
                  BO and BOGP: $E_\textrm{lowest} = 1507.1 ~ \textrm{cm}^{-1}$).
           	 The energy of the light-induced conical intersection (LICI) is $29390.5 ~ \textrm{cm}^{-1}$ referenced to $E_\textrm{lowest}$.
           	 The cavity wavenumber and coupling strength are 
		     $\omega_\textrm{c} = 30245.5 ~ \textrm{cm}^{-1}$ and $g = 0.1 ~ \textrm{au}$, respectively.}
  \label{tbl:levels}
  \begin{tabular}{ccc}
    \hline
    Eigenstate label & $~~(E-E_\textrm{lowest}) ~ / ~ \textrm{cm}^{-1}$ & Photonic part population \\
    \hline
    $0$   & $28863.7$ & $0.11$ \\
    $1$   & $29805.9$ & $0.46$ \\
    $2$   & $30039.7$ & $0.12$ \\
    $3$   & $30620.1$ & $0.37$ \\
    \hline
    $1\textrm{A}$   & $29523.7$ & $0.24$ \\
    $1\textrm{B}$   & $29924.0$ & $0.33$ \\
    $3\textrm{A}$   & $30599.5$ & $0.07$ \\
    $3\textrm{B}$   & $30669.0$ & $0.33$ \\
    $4\textrm{B}$   & $31200.7$ & $0.30$ \\
    \hline
    $0\textrm{b}$   & $28819.7$ & $0.12$ \\
    $1\textrm{a}$   & $29699.1$ & $0.15$ \\
    $1\textrm{b}$   & $29766.2$ & $0.31$ \\
    $2\textrm{a}$   & $30039.3$ & $0.14$ \\
    $2\textrm{b}$   & $30063.7$ & $0.32$ \\
    $3\textrm{a}$   & $30627.2$ & $0.05$ \\
    $3\textrm{b}$   & $30713.2$ & $0.37$ \\
    \hline
  \end{tabular}
\end{table}

\end{document}